\documentclass[english,notitlepage, twocolumn]{revtex4-1}
\usepackage[T1]{fontenc}
\usepackage[latin9]{inputenc}
\usepackage{color}
\usepackage{amsmath}
\usepackage{amssymb}
\usepackage{graphicx}

\usepackage{babel}
\begin{document}

\title{Non-line-of-sight Imaging}

\author{D. Faccio$^1$, A. Velten$^2$, G. Wetzstein$^3$}

\affiliation{$^1$School of Physics \& Astronomy, University of Glasgow, G12 8LT Glasgow, United Kingdom\\
$^{2}$Department of Biostatistics and Medical Informatics, University of Wisconsin, Madison, WI 53706, USA.\\
$^{3}$Department of Electrical Engineering, Stanford University, Stanford, CA 94305, USA}

\email{daniele.faccio@glasgow.ac.uk, velten@wisc.edu, gordon.wetzstein@stanford.edu}

\begin{abstract}
%
Emerging single-photon-sensitive sensors combined with advanced inverse methods to process picosecond-accurate time-stamped photon counts have given rise to unprecedented imaging capabilities. Rather than imaging photons that travel along direct paths from a source to an object and back to the detector, non-line-of-sight (NLOS) imaging approaches analyze photons  {scattered from multiple surfaces that travel} along indirect light paths to estimate 3D images of scenes outside the direct line of sight of a camera, hidden by a wall or other obstacles. Here we review recent advances in the field of NLOS imaging, discussing how to see around corners and future prospects for the field. 
\end{abstract}
\maketitle



\begin{figure*}[t]
\centering
\includegraphics[width=\textwidth]{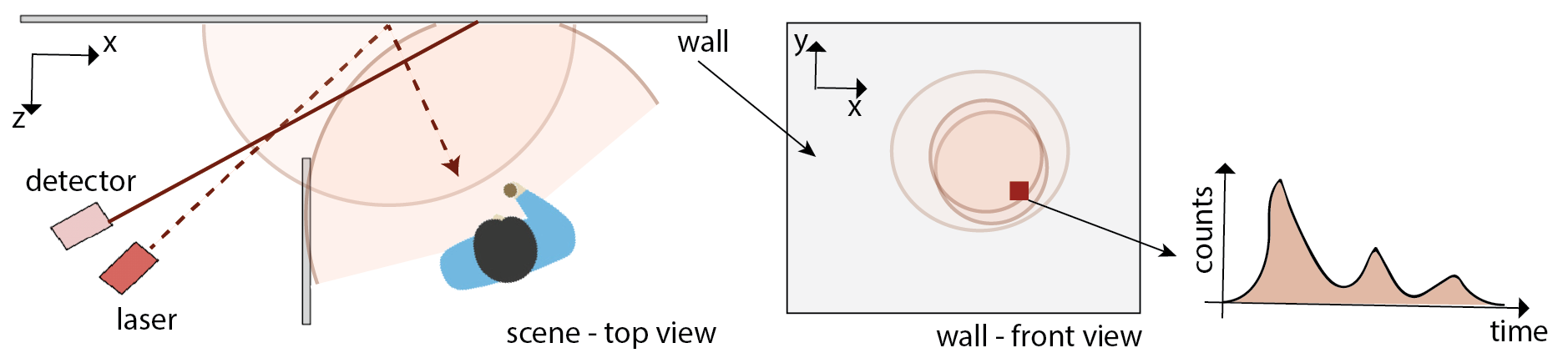}
\caption{Schematic layout of time-resolved NLOS imaging. A visible wall is illuminated with a pulsed laser source. Light scattered from this wall extends into the obscured region and indirectly illuminates hidden objects, which in turn scatter the light back to the wall where it is recorded by a time-resolved detector, typically a SPAD sensor. The middle panel gives a schematic view of how the waves scattered from the hidden objection may look like on the observation wall when ``frozen'' at a specific time. The ellipsoids observed on the wall are the intersection with the wall of spherical scattered waves from the object. The far-right graphs shows a schematic example of the temporal trace of photon counts observed at a given pixel on the wall. The various peaks correspond to the scattered spherical waves expanding outwards as time evolves.}
\label{fig:ToF_layout}
\end{figure*}

{\bf{Introduction.}}  
Imaging objects outside a camera's direct line of sight is of fundamental importance for applications in robotic vision, remote sensing, medical imaging, autonomous driving, and many other domains. For example, the ability to see hidden obstacles could provide autonomous vehicles with a way to avoid collisions, drive more efficiently, and plan driving actions further in advance. LiDAR and other 3D imaging systems commonly used in automotive sensing measure the time it takes a light pulse to travel along a direct path from a source, to a visible object, and back to a sensor. Non-line-of-sight (NLOS) imaging goes one step further by analyzing {light scattered from multiple surfaces} along indirect paths with the goal of revealing the 3D shape and visual appearance of objects outside the direct line of sight~\cite{kirmani2009,velten}  (see Fig.~\ref{fig:ToF_layout}).

NLOS imaging poses several challenges. First, only a few of the many recorded photons carry the information necessary to estimate hidden objects. Whereas the photon count of light directly reflected from {a single scattering point}  falls off with a factor proportional to the inverse of the square distance, the signal strength of   {multiply} scattered light decreases several orders of magnitude faster. 
Robustly detecting and time-stamping the few   {indirectly scattered} photons in the presence of the much brighter signal returning directly from the visible scene requires high dynamic range or gated single-photon-sensitive detectors. 
Second, the inverse problem of estimating 3D shape and appearance of hidden objects from intensity measurements alone is ill-posed. Advanced imaging systems that leverage picosecond-accurate time-resolved measurement capabilities, mathematical priors on the imaged scenes, or other unconventional approaches are required to solve the NLOS problem robustly. Third, the inverse problems associated with NLOS imaging are extremely large. Developing efficient algorithms to compute solutions in reasonable time frames and using available memory resources is crucial to make this emerging imaging modality practical.

Over the last few years, a variety of different approaches to addressing the NLOS problem have been proposed. Some of these focus on advanced measurement systems, using femtosecond and picosecond time-resolved detectors~\cite{velten,o2018confocal,liu_non-line--sight_2019,Lindell:2019:Wave}, interferometry~\cite{Jacopo,Ori}, acoustic systems~\cite{Lindell:2019:Acoustic}, passive imaging systems~\cite{bouman2017,saunders2019computational,boger2018}, or thermal imaging system~\cite{maeda:2019,thermal}, while others explore various models of light transport that make certain assumptions on reflectance or other properties of the hidden scenes. At the convergence of physics, signal processing, optics, and electronics, NLOS imaging is an interdisciplinary challenge that has seen much progress over the last few years. Nevertheless, continued effort on both theory and experimental systems is necessary to make the idea of seeing around corners practical ``in the wild''.

In this perspective, we discuss the emerging field of NLOS imaging and aim at making it accessible to the reader by categorizing existing approaches by the types of measurement systems they use and also by their algorithmic approaches. Time-resolved imaging systems leveraging pulsed light sources along with single-photon detectors are  highlighted as one of the most promising directions towards practical solutions in this area. Although 3D imaging~\cite{Altmanneaat2298}, or direct ranging and imaging through scattering media are problems closely related to NLOS imaging, these will not be covered in depth. 




To illustrate the measurement process of time-resolved NLOS imaging systems, Figure~\ref{fig:ToF_layout} shows an example scene observed from a top view where a pulsed laser with a pulse width,   {for example} in the range 100~fs--100~ps, illuminates a wall at one point. The light reaching the wall subsequently scatters into the hidden region where it re-scatters off of any hidden objects before returning to the wall where the time-resolved indirect light transport is measured. The front view of all transverse measurements taken on the wall shows a schematic representation of what these re-scattered waves, originating from the hidden objects might look like if observed frozen at a given time: individual areas on the object scatter back spherical waves, which upon intersecting the wall, give rise to ellipsoids that expand outwards in time. It is these time-varying ellipsoids that contain all the information required to then reconstruct a full 3D image of the hidden scene. The key requirement for time-resolved NLOS approaches is the temporal resolution of the detector---it must be sufficiently high to freeze light in motion \cite{RoPP_review}. Light travels $\sim3$~cm in 100~ps, which motivates the desired temporal resolution of the imaging system as this will map directly onto the achievable transverse and axial resolution of estimated 3D images.

\begin{figure}[t]
\includegraphics[width=8cm]{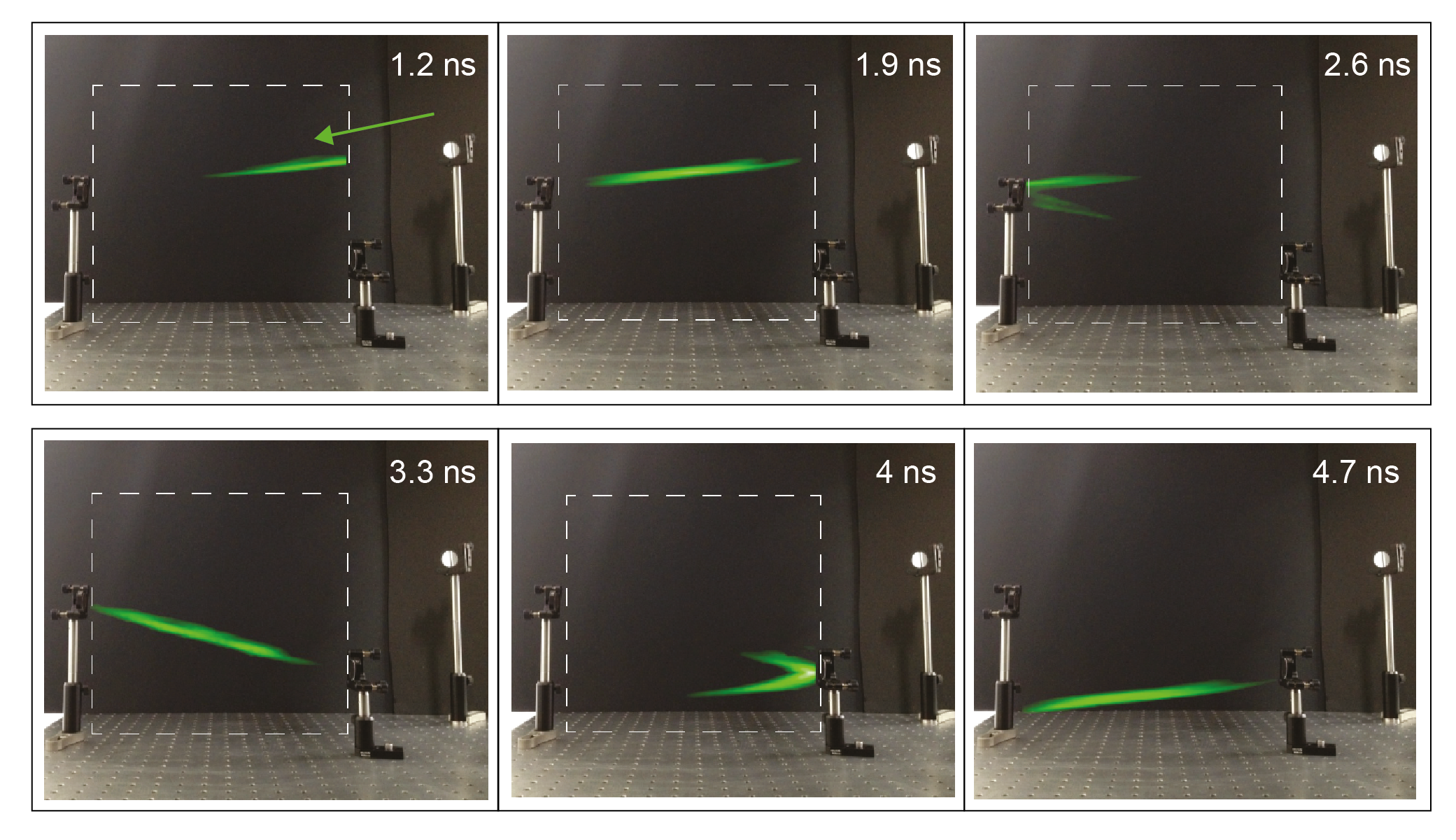}
\begin{center}
\caption{Many NLOS imaging approaches build on time-resolved measurements of light transport. The capability of recording light in flight at picosecond timescales is demonstrated here for a pulse of light propagating between three mirrors. The laser first hits the mirror on the right and is directed towards the field of view of the SPAD camera, as indicated by the green arrow in the first frame. The FOV is represented by dashed rectangles and corresponds to a $35 \times 35$~cm$^2$ region. In the first and second frames, we show the laser pulse entering the FOV. In the second, third and fourth frames, we see the light being reflected by the mirrors, before exiting the FOV in the last frame. Figure adapted from~\cite{me1}.}
\label{fig:transientimaging}
\end{center}
\end{figure}

\begin{figure}[!t]
\centering
\includegraphics[width=8cm]{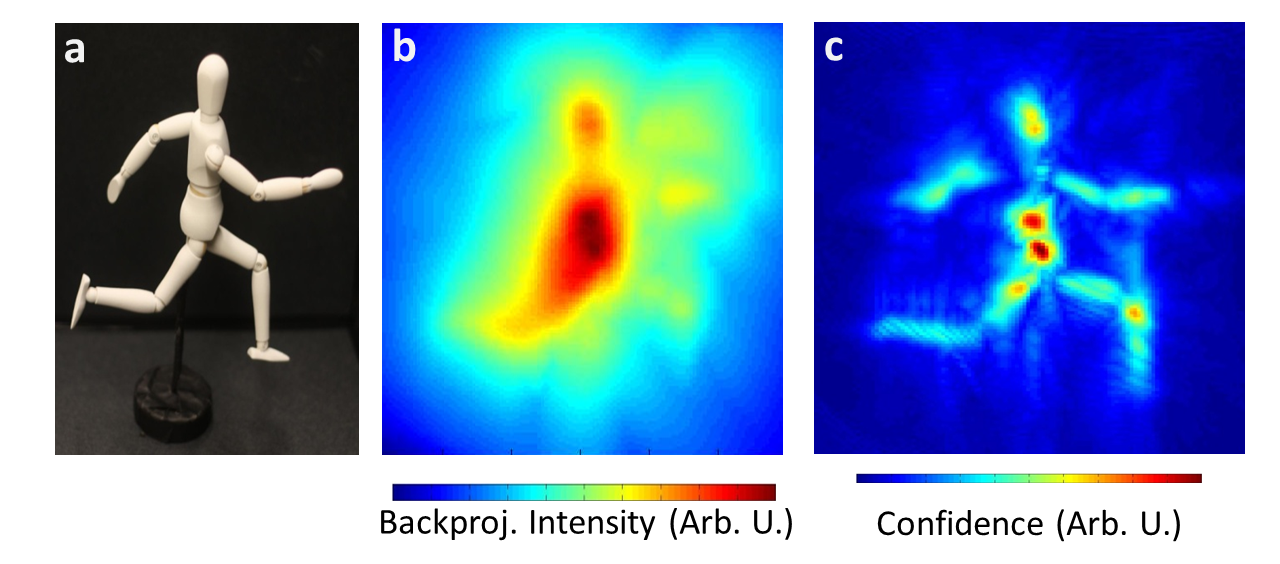} 
\caption{First experimental demonstration of ``looking around corners''. A mannequin behind a corner (a) is recovered from time-resolved measurements using unfiltered (b) and filtered (c) backprojection algorithms. Figure adapted from~\cite{velten}.}
\label{fig:NLOS_result}
\end{figure}

{\bf{Imaging at the Speed of Light.}}
The concept of freezing light in motion, sometimes referred to as ``light-in-flight'' or ``transient'' imaging, is not specific to NLOS imaging \cite{RoPP_review}. Several techniques for light-in-flight imaging have been proposed starting from the 1960's when researchers at Bell Labs used nonlinear optical gating techniques to create an ultrafast shutter, therefore  extending the basic concept of the mechanical shutter used in many high-speed cameras to that of a shutter that is activated by light itself. Another ingenious approach that effectively paved the way for true light-in-flight imaging was developed by N. Abramson in the '70s and relies on standard holographic techniques \cite{holo} where the reference field is now a laser pulse that is spatially extended and hits the photographic plate at an angle \cite{abramson_light--flight_1978,6HR1983,Abramson:85,abramson}. The result is a hologram where different transverse locations on the exposed photographic plate correspond to  different times in the scene due to the different arrival times of the tilted reference pulse. Viewing the photographic plate at different lateral positions provides an image at different times with resolutions in the order of picoseconds or even less. A related technique was introduced based on a generalisation of optical coherence tomography  that also relies on interference of the light reflected from a scene or object with a reference field. Reconstruction of transient light scenes with very high (tens of microns) spatial resolution with 15 trillion frames per second are obtained through detection of interference fringes as the interferometer delay is varied \cite{gkioulekas2015micron}.\\
Despite the success of these and related approaches, they all rely in one form or another on direct control over the scene itself, for example  through the requirement of knowledge of the precise temporal arrivals from the scene (in order to synchronise the reference optical pulse) or through careful tailoring of an object or its position that ultimately results in limitations in the size or complexity of the scene. The key to technological applications of light-in-flight imaging and in particular to NLOS imaging was the development of camera technology that operates on the basis of in-built electronic gating and synchronisation. Examples of such cameras are time-of-flight cameras, streak cameras, intensified CCDs and single-photon-avalanche-diode (SPAD) arrays.\\
Transient imaging using time-of-flight (ToF) cameras provides a 3D image of a scene that can also be applied to NLOS ~\cite{18ToFMIT2013,Heide1,peters2015solving} and offers the distinct advantage of being very low-budget with commercial ToF cameras costing in the order of \$100. These cameras illuminate the scene with a sinusoidal-modulated (typically 10-100 MHz or higher) light beam. The return signal is demodulated against a reference sine wave from which a phase delay is extracted that is directly related to the time-of-flight and hence to the propagation distance within the scene ( {see e.g. \cite{jarabo2017recent,RoPP_review} for an overview}). \\
Higher temporal resolution and better light sensitivity, both key parameters for NLOS, can be obtained with more complex and expensive cameras. 
For example, the first  demonstration of full 3D NLOS imaging was performed using a streak camera to obtain precise reconstruction of a small mannequin object~\cite{velten}, as shown in Fig.~\ref{fig:NLOS_result}. These cameras rely on a photocathode to convert the incoming photons into electrons. 
The electrons can then be `streaked' by a time-varying electric field, therefore mapping time onto transverse position. The streaked electrons are detected on a standard CCD camera after re-conversion back to photons on a phosphor screen. The use of one spatial dimension for the temporal streaking implies that these cameras can only see one line of the scene at a time,  {a limitation that can be offset for NLOS imaging by instead scanning the illumination laser spot \cite{velten,velten_femto-photography:_2013}}. Techniques have been implemented that allow to fully open the input slit and, by computational fusion with data from a CCD, obtain a full 2D image without any need for scanning~\cite{12SSN2014,15RUI2016,CUP2}. Interestingly, these full-imaging approaches have not been applied yet to NLOS imaging. \\
An alternative approach is based on the use of intensified CCD cameras (iCCD). iCCDs rely on a microchannel plate that is  electronically gated so that electrons generated by an input photocathode are amplified only for a short gate-time before being re-converted back to light on a phosphor screen and detected on a CCD or CMOS camera. Typical gate times are of order of nanoseconds but can be as short as 100 ps or less. As for all of the imaging techniques reviewed here, iCCDs can also be used for NLOS imaging \cite{Velten_gated}.\\
These and subsequent techniques applied to light-in-flight imaging have sufficient precision for example to observe distortions in the final video due to the finite speed of light, such as apparently inverted motion of refracted waves from a bottle or apparent superluminal motion of light pulses \cite{jarabo_relativistic_2015,LaurenzisOL}. 
A 100 ps-gate iCCD was used for example to record the apparent time reversal of events occurring during light propagation: the intersection of a plane wave hitting a wall at an angle $\theta$  will travel at speed $c/\sin\theta$ and is therefore always superluminal. The transient imaging of the  scattering of light from this intersection plane on the wall will reveal an apparent motion in the opposite direction to that actually followed by the light pulse \cite{clerici_super} in much the same way that a piece of music played by a speaker moving faster than the speed of sound will be heard backwards \cite{rayleigh}. \\
Moving beyond the first 3D NLOS imaging based on streak cameras~\cite{velten}, work ensued to improve upon some of the limitations encountered in these measurements that required several hours of data acquisition with particular emphasis on improving acquisition speed (is video frame rate imaging possible?), light sensitivity (can we extend the observation area to entire rooms and observe human-sized objects?), portability (can we deploy this technology in the real-world?) and cost (is there a technology that does all the above with similar costs to a ToF camera?). Single photon avalanche diode (SPAD) detectors appear to address most if not all of these points.\\
SPADs are semiconductor structures similar to  {avalanche photodiodies}, APDs. A photodiode with a large bias voltage results in carrier multiplication such that the absorption of a single photon causes a breakdown that can be detected and processed by external electronics. Time-to-digital converters measure the time between the emission of an illumination pulse and the detection of an associated returned photon on the SPAD. A time-correlated single-photon counter (TCSPC) is then used to form a histogram of photon arrival times \cite{TCSPC}. SPADs achieve single photon sensitivity with  {photon detection efficiencies} up to 40\% and exceptionally low dark count rates  down to 1-10 photons per second in the visible spectrum. After the detection of a photon the detector is blind for a hold-off period (dead time) of 10s to 100s of nanoseconds, thus limiting the achievable maximum count rate. The histogram will give a precise measurement of the light pulse temporal profile as long as the measurement is performed in a photon-sparse regime, i.e. a regime in which the likelihood of more than one photon hitting the detector during the dead time (referred to as pile-up) is significantly less than one. Accounting for the SPAD dead-time, this provides a maximum allowed count rate, in order to avoid photon pile-up distortion effects, of the order of 1-10 MHz. \\
SPAD detectors are available in both single pixel and arrayed (camera) format at both visible \cite{RH2,RH3,RH4,RH5,Tosi2,Tosi3,Cammi,Tosi4,RH1,Tosi1,linospad,RH-vertical} and infrared wavelengths \cite{itzler1,itzler2,itzler3}.  SPAD cameras have been employed for light-in-flight imaging, where the single-photon sensitivity allowed to capture a light pulse propagating in free space where photons collected on the camera originated from Rayleigh scattering in air (as opposed to scattering from a surface or enhanced scattering in a diffusive medium) \cite{me1}. A selection of frames from the full video are shown in Fig.~\ref{fig:transientimaging}: the 32x32 pixel SPAD camera had a temporal resolution of $\sim50$ ps corresponding to 200 million frames/second. Although not as fast as some of the techniques discussed above that can reach above a trillion frames/second, this is still sufficient to freeze light in motion with a blur of only 1.5 cm. This minor loss of temporal resolution comes at the benefit of compactness and ease of use (the camera is based on standard CMOS technology, is commercially available \cite{PF,MPD} and is small enough to be integrated into a smartphone \cite{LG}), high data acquisition rates (NLOS data acquisition has been demonstrated with sub-second timescales \cite{single_pixel_NLOS}) and with interference filters at the specific laser illumination wavelength can also be deployed outdoors and in daylight conditions \cite{SusanFar,o2018confocal}. Recent advances have shown video frame rate acquisition of transient images using SPADs \cite{lindell2018towards} also in more standard LIDAR configurations deployed outdoors over kilometer distances  \cite{buller2017}. \\
The first application of SPAD array sensors to NLOS imaging was in a simpler configuration where only the position of the target was assessed rather than its full 3D shape. This allowed acquisition and processing times of the order of 1 second for a moving target, both in a small-scale laboratory setup \cite{me2} but also for detecting people behind a corner on larger scales (>50 m distance from the detector) \cite{SusanFar}.  Single pixel gated SPADs \cite{buttafava2015} and line arrays \cite{wetzstein1} with a scanning laser spot have also been used to acquire full 3D scenes and currently appear to be one of the preferred approaches for NLOS imaging with most results over the past few years employing SPADs either in single pixel or array format. \\
 {The temporal resolution actually required from the detector will depend on a number of factors, including the illumination pulse length and the task at hand. For example, for transient imaging such as capturing a light pulse in flight, there is no need to employ a detector with temporal resolution shorter than the pulse length. For 100 ps or longer pulses, this can easily be achieved with the techniques described above. For femtosecond pulses, such as those available from standard femtosecond oscillators, the current resolution of detectors, limited to ten or more picoseconds will unavoidably result in temporal blur of the pulse that will be of order of 0.3-1 cm, compared to the 30 $\mu$m of a 100 fs pulse. However, when considering NLOS imaging, the detector temporal resolution will directly affect both transverse and depth resolution of the 3D image reconstruction, as discussed below, Eq.~(\ref{eq:resolution}).}\\
Looking forward, most current SPAD arrays are developed for LiDAR imaging. NLOS applications require higher temporal resolutions, better fill factors, the ability to gate out direct light from the relay surface and a more flexible way to read out photon time stamps from those SPAD pixels that see a photon. There is therefore a need for SPAD arrays specifically designed with NLOS applications in mind.








{\bf{Inverse Light Transport for Time-resolved NLOS Imaging.}}

In this section we outline a general image formation model for time-resolved NLOS imaging, overview recently proposed inverse methods for this imaging modality, and discuss fundamental bounds on achievable resolution of the hidden object reconstructions. 

\emph{Image Formation Model.}
A time-resolved detector, such as a SPAD, measures the incident photon flux as a function of time relative to an emitted light pulse. The detector is therefore used to record the temporal impulse response of a scene, including direct and global illumination, at sampling positions $x',y'$ on a visible surface (see Fig.~\ref{fig:ToF_layout}), resulting in a 3D space-time volume that we refer to as the \emph{transient image} $\tau$. As discussed in the previous section, a transient image contains both directly reflected photons but also photons that travel along indirect light paths. The direct illumination, i.e., light emitted by the source and scattered back to the detector from an object, contains all information necessary to recover the shape and reflectance of visible scene parts. This is commonly done for 3D imaging and LiDAR. For NLOS imaging, the direct light is typically not considered because it does not contain useful information of the hidden scene and it can be easily removed from measurements (for example by using the fact that it arrives earlier than multiple-surface reflected photons and can therefore be gated out).
The image formation model for the time-resolved indirect light transport of a confocal NLOS system~\cite{o2018confocal}, i.e. one in which both the laser illumination and the subsequent detection are at the same point $x',y'$ on the visible surface, can be formulated as
\  {
\begin{align}\label{eq:imageformation}
	\tau \left( x', y', t \right) & = \int \!\!\!\!  \int \!\!\!\! \int_{\Omega} \,\, \frac{1}{r^4} \,\, \rho \left( x,y,z \right) g \left(x',y',x,y,z \right)\,\, \\
	& \delta \left( 2\sqrt{\left( x' - x \right)^2 + \left( y' - y \right)^2 + z^2} - tc \right)   \textrm{d}x \textrm{d}y \textrm{d}z, \nonumber
\end{align}
}
where the Dirac delta function $\delta\left( \cdot \right)$ relates the time of flight $t$ to the distance function $r = \sqrt{ \left( x' - x \right)^2 + \left( y' - y \right)^2 + z^2 } = tc / 2$. Here, $c$ is the speed of light and $x,y,z$ are the spatial coordinates of the hidden volume. For convenience, we assume that the sampling locations $x',y'$ are located on the plane $z=0$, \  {that the laser pulse is infinitesimally short, and we only consider indirect light transport that bounced precisely three times after emission by a light source and before being detected: off of a visible surface within the line of sight, then off of a hidden surface outside the line of sight, and finally, once more off of the visible surface again. The function $g$ absorbs miscellaneous time-independent attenuation effects that depend on the hidden surface normals, reflectance properties of the hidden scene, visibility of a hidden point from some sampling point $x',y'$, and several other factors.} Note that each measurement in the confocal configuration integrates over spherical surfaces in the hidden scene. More general non-confocal configurations are also common, where the detector samples the time-resolved indirect light transport at one point on the wall while the laser directly illuminates a different point on the visible surface~\cite{velten,liu_non-line--sight_2019}. The laser point or the detection point can then be scanned independently from each other. In this more general configuration, measurements integrate along elliptical surfaces. \  {Moreover, higher-order bounces of indirect light transport could also be considered to model indirect reflections of light within the hidden scene, although these become increasingly difficult to measure.}


This common image formation model is at the core of most NLOS imaging approaches. \  {The effects modeled by $g$ make this a nonlinear image formation model. Several approaches reported in the literature work with a linearized approximation of Equation~\ref{eq:imageformation}, where $g=1$. This linear approximation is easier to invert than the nonlinear model but it makes several additional assumptions on the light transport in the hidden scene: light scatters isotropically and no occlusions occur between different scene parts outside the line of sight. \  {It is important to note, that line of sight imaging problems are made nonlinear in a similar fashion if surface normals, BRDFs, and occlusions are included in the model. This is why, typically, line of sight imaging systems operate on linearized transport models as well.} Various approaches to solving both the linearized and nonlinear NLOS problem are discussed in the following. The former approach reduces to approximating} or solving the large, linear equation system $\boldsymbol{\tau} = \boldsymbol{A} \boldsymbol{\rho}$, where $\boldsymbol{\tau}$ represents the discretized transient measurements, $\boldsymbol{\rho}$ are the unknown reflectance values of the hidden scene albedo, and $\boldsymbol{A}$ is a matrix describing the indirect time-resolved light transport.



\emph{\  {Heuristic Solutions}} for estimating the shape and reflectance of the hidden volume have been very popular. One of the most intuitive of these approaches is to relate the measured times of the first-returning indirect photons and relate these to the convex hull of the hidden object or scene~\cite{tsai2017geometry}. Alternatively, simple parametric planar models can be fitted to represent the hidden scene~\cite{pediredla2017reconstructing}. Another area still in its infancy is the utilization of active capture methods that shape illumination and detection to optimize capture based on the anticipated scene content. Spatial refocusing after the first scattering surface can be controlled using spatial-light-modulators and the focused spot can be scanned across the scene \cite{ilya}. Temporal focusing uses an illumination pulse that is shaped in space and time to create an illumination pulse at an area in the hidden scene~\cite{pediredla2019snlos}. These techniques can improve the signal to noise ratio and resolution for the obtained reconstruction.

\emph{Backprojection Methods} are some of the most popular methods for NLOS image reconstruction, which approximate the hidden volume as $\boldsymbol{A}^T \boldsymbol{\tau}$ and optionally apply a filtering or other post-processing step to this result (Fig.~\ref{fig:NLOS_result}). Similar strategies are standard practice for solving large-scale inverse problems, for example in medical imaging. Indeed, the inverse problem of confocal NLOS scanning approaches are closely related to the spherical Radon transform~\cite{Tasinkevych2015CircularRT}, whereas the general non-confocal scanning approach is similar to the elliptical Radon transform~\cite{moon2014}. Filtered backprojection methods are standard solutions to these inverse problems. Both computational time and memory requirements of these Radon transforms are tractable even for large-scale inverse problems. Hence, several variants of backprojection algorithms have been explored for NLOS imaging~\cite{5MIT2012,6HSO2012,buttafava_time-gated_2014,laurenzis_feature_2014,arellano_fast_2017}, \  {although these algorithms have a computational complexity of $O(N^5)$ for a total of $N$ voxels}. Similar to limited-baseline tomography problems~\cite{kak2002principles}, non-line-of-sight problems are typically ill-posed inverse problems because acquired measurements usually do not sample all Fourier coefficients. In microscopy and medical imaging, this is known as the ``missing cone'' problem. To estimate these missing components, statistical priors need to be incorporated into the inverse method to fill in these parts using iterative solvers.

\emph{Linear Inverse Methods} have been proposed to solve the convex optimization problem of estimating $\boldsymbol{\rho}$ from $\boldsymbol{\tau}$. Several of these approaches aim at using iterative optimization methods to solve this problem~\cite{6HSO2012,wu_frequency_2012,Heide2}, but these are typically very slow. The light cone transform was proposed as a closed-form solution to the linear inverse problem and it efficiently solves the exact linear inverse problem assuming a smoothness prior on the reconstructed volume \  {with a computational complexity of $O(N^3 log N)$}~\cite{o2018confocal}. An implementation of this method on graphics processing units was demonstrated to achieve real-time reconstruction rates~\cite{OToole:2018:realtime}.
\begin{figure*}[t]
\centering
\includegraphics[width=\textwidth]{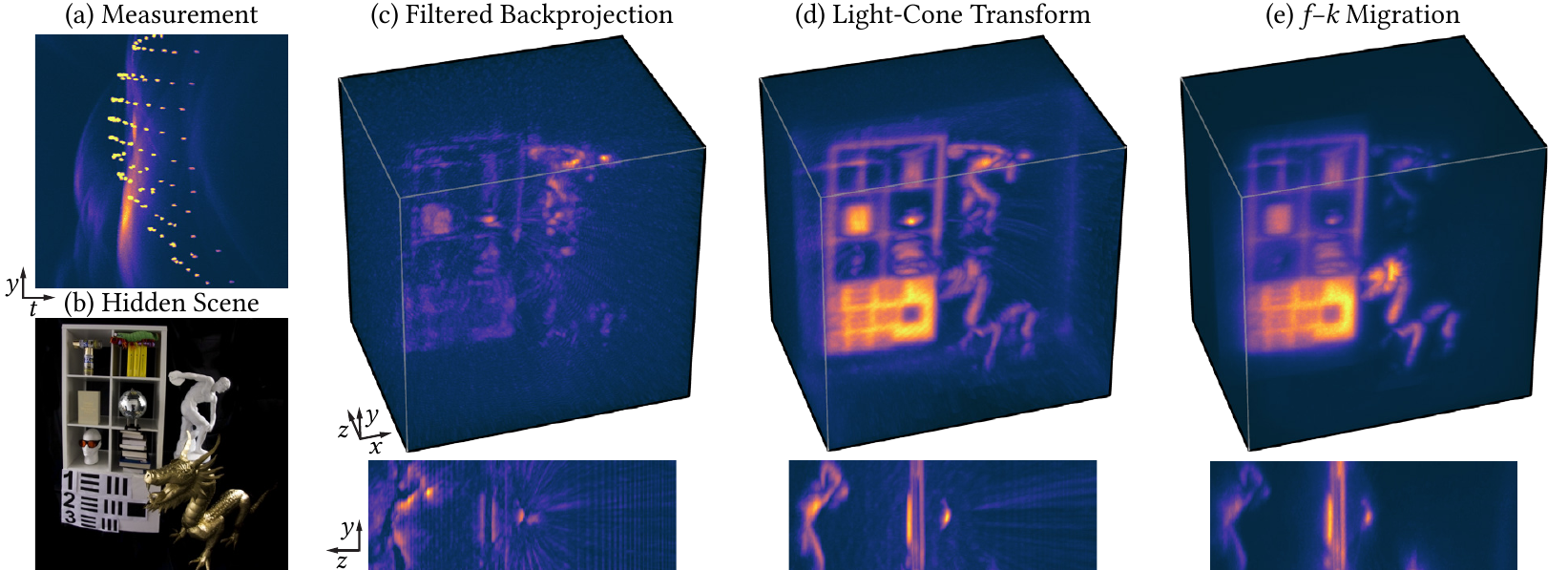}
\caption{
Non-line-of-sight (NLOS) reconstructions of a hidden, room-sized scene. (a-b) One approach to NLOS imaging is to capture time-resolved measurements
sampled across a visible surface, and reconstruct the 3D shape and reflectance of the hidden scene. A disco ball produces the bright dots seen in the measurements of indirect light transport (a), and other diffuse and glossy objects produce the streaks. (c) Among the various methods for reconstructing shape and reflectance from these measurements, filtered backprojection (FBP) is conceptually one of the simpler methods; it involves a delay-and-sum (backprojection) operation of the time-resolved measurements, followed by a heuristic high-pass filter on the result. (d) The light-cone transform (LCT) is a fast reconstruction algorithm that produces more accurate reconstructions in less time, but it requires the hidden objects to be either diffuse or retroreflective. (e) NLOS imaging with f-k migration is both fast and versatile. The wave-based nature of this inverse method is unique in being robust to objects with diverse and complex reflectance properties, such as the glossy dragon, the diffuse statue, and the reflective disco ball shown in this scene. All volumes are rendered as maximum intensity projections.  Figure adapted from~\cite{Lindell:2019:Wave}.}
\label{fig:NLOS_result2}
\end{figure*}
\begin{figure*}[t]
\centering
\includegraphics[width=\textwidth]{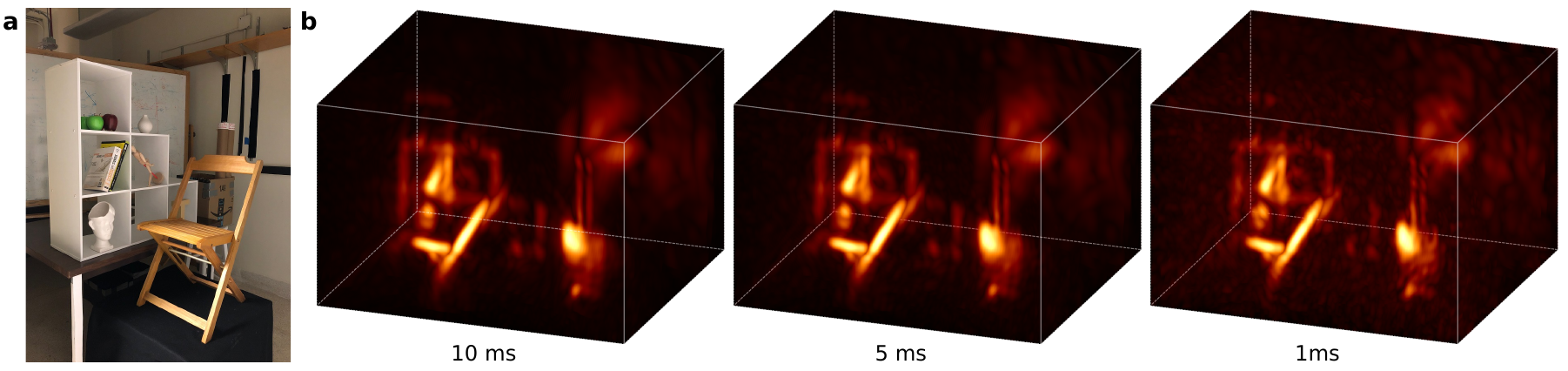}
\caption{Reconstructions of a large scene using the Phasor Field Virtual Wave approach. The data is collected with a single pixel SPAD using point scanning to emulate a large detector array. The exposure time per scanned point is shown under the Figures. The entire scan involves 24000 points leading to data collection times of 4 minutes, 2 minutes, and 24 seconds respectively. The scene is about 2 meters wide and 3 meters deep. Figure adapted from~\cite{liu_non-line--sight_2019}}
\label{fig:phasor_result}
\end{figure*}


\emph{Inverse Light Transport with Partial Occlusions, Surfaces, and Normals} has received much attention in recent research proposals, because some of the simplifying assumptions of the image formation model (Eq.~\ref{eq:imageformation}) can be lifted \  {by solving the nonlinear problem rather than a linearized approximation}. For example, several time-resolved methods have included partial occlusions within the hidden scene in the image formation model~\cite{heide2019non,thrampoulidis2018exploiting,xu2018revealing}. Interestingly, it has been shown that occlusions and shadows in the hidden can also be exploited to facilitate passive NLOS approaches that do not require time-resolved imaging systems~\cite{bouman2017,seidel2019corner,baradad2018,saunders2019computational}. However, the associated inverse problems are much more ill-posed than with active imaging and the proposed algorithms often make restrictive assumptions. A few recent approaches have also incorporated hidden surface normals into the image formation model~\cite{heide2019non,xin2019theory}, which can further help improve reconstruction quality. Finally, an emerging research direction is to aim at reconstructing hidden surfaces, rather than volumes, directly from the transient measurements~\cite{iseringhausen2018,xin2019theory,tsai2019beyond,Young:2020}. High-resolution volumes are memory-inefficient data structures and can quickly exceed available computational resources. Therefore, a trade-off between level of detail of a reconstructed volume and memory requirement may have to be made in practice. Surface representations have the potential to represent finer geometric detail with fewer computational resources. It still seems unclear, however, what the ``best'' representation for general NLOS imaging is.

\emph{Wave Optics Models}, rather than the above outlined geometric optics model, have recently been explored for transient imaging configurations with time-resolved detectors and pulsed light sources~\cite{reza2018physical,liu_non-line--sight_2019,teichman2019,reza2019experimemts_phasor,Dove_19,Lindell:2019:Wave} (see Figs.~\ref{fig:NLOS_result2},\ref{fig:phasor_result}). In these methods, the light transport in the hidden scene is modeled using the time-dependent wave equation or other models from physical optics. A similar concept was also applied to NLOS data captured in the Fourier domain by an amplitude-modulated continuous-wave light source~\cite{Raskar_PMD}.

\  {The algorithms in this category do not necessarily try to solve the inverse problem of estimating the hidden geometry directly, as most of the methods discussed above do. Rather, the transient image is treated as a virtual wave field and propagated backwards in time to a specific time instant. The geometry estimation problem then becomes that of relating the hidden geometry to specific properties of the temporally evolving wave field. Just like in a line of sight camera, the problem is thus divided into a linear operator that estimates the wave in the hidden scene (i.e., the image) and a nonlinear problem of estimating geometry, BRDF, etc. from the image.}

There are several benefits when considering a wave optics model for the NLOS problem. First, some of these approaches have been experimentally shown to be more robust to different types of reflectance properties of the hidden surfaces. For example, glossy, specular, diffuse, or retro-reflective materials can all be treated with the same method whereas geometric optics approaches either \  {have to know and model the reflectance properties a priori or estimate them along with the hidden geometry.} Second, wave models make it easier to draw the connection between NLOS imaging and related work in areas such as radar, seismic imaging, ultrasonic imaging, and other established fields. \  {For example, it was recently shown how range migration techniques originally developed in the seismic imaging community~\cite{stolt1978,margrave2018}, and later used synthetic aperture
sonar (SAS)~\cite{callow2003,sheriff1992}, ultrasound imaging~\cite{garcia2013}, and synthetic aperture radar~\cite{cafforio1991}, result in some of the fastest and most robust NLOS imaging techniques~\cite{Lindell:2019:Wave}.} It should be noted that the phase information of the light wave used in these experiments is not measured or required. What is used is the phase and wavefront of an intensity wave riding on the optical carrier wave. The phase of this wave is related to the time of arrival of the signal photons, not to their optical phase. The phase of the light wave is typically not accessible with time-resolved NLOS imaging systems. The time-of-flight information of indirect light transport must instead be used to estimate object shape, which makes the associated inverse problems different.\\


\emph{Data-driven Approaches} are an emerging tool for NLOS reconstructions. Neural Networks can reconstruct hidden scenes from steady state data captured with a continuous light source and a conventional camera~\cite{Tancik:18,chen2019steady}. However, practical application of neural networks to time-of-flight data faces the difficulty to generate sufficient amounts of training data, although training on real people and classification of a small set of individuals and of their positions has been shown~\cite{pier}. 


\emph{NLOS Tracking} of objects and people with time-resolved imaging systems is also an active area of research~\cite{velten_motion,SusanBox,SusanFar,gariepy_tracking_2015,o2018confocal}. The tracking problem is significantly simpler than reconstructing a full hidden 3D volume, which makes it computationally more efficient to implement. These approaches pave the way for future research that goes beyond hidden shape reconstruction and that could aim at classification \cite{pier}, object detection, target identification, or other inverse problems that build on transient light transport.


\  {\emph{NLOS Imaging without a Relay Wall} is another emerging paradigm in this  area. Most existing NLOS approaches require the imaging system to scan a large area on a visible surface, where the indirect light paths of hidden objects are sampled. In many applications, however, optical access to a large scanning area may not be available. By exploiting scene motion, one can derive inverse methods that estimate both the hidden object's shape and unknown motion trajectory simultaneously from transient images~\cite{metzler2019keyhole}. This is a significantly more challenging and ill-posed problem than conventional NLOS imaging because the light transport is only measured along a single optical path, but it may further extend the application space of NLOS imaging techniques.}


\emph{Resolution Limits.}
The resolving power of conventional, diffraction-limited imaging systems is fundamentally limited by the numerical aperture of the optics and the wavelength at which they operate~\cite{born2013principles}. Time-resolved NLOS imaging also obeys fundamental resolution limits. These are primarily defined by two factors: (i) the area on the visible surface over which the  {time-resolved indirect light transport of} the hidden scene is recorded and (ii) the temporal resolution of the imaging system. The first factor, the scanning area, is analogous to the numerical aperture of a conventional imaging system---the larger the scanning area or numerical aperture, the better the transverse resolution. The second factor, temporal resolution, is somewhat analogous to the wavelength limiting resolution of conventional systems. Together, these two characteristics of an NLOS imaging system define both transverse and axial resolution of a hidden volume that can be estimated unambiguously, i.e. without the use of statistical priors.

Formally, we define the resolution of a NLOS system as the minimum resolvable distance of two scatterers that can be resolved. These two scattering points are resolvable in a hidden 3D space only if the measurements of their indirect reflections are resolvable in time. Assuming that the temporal resolution of the system is given by the full width at half maximum (FWHM) of its temporal impulse response, transverse and axial resolutions are
\begin{equation}
	\Delta x \geq \frac{c \cdot \sqrt{w^2 + z^2}}{2 w} \textrm{FWHM}, \quad \Delta z \geq \frac{c \cdot \textrm{FWHM}}{2}.
	\label{eq:resolution}
\end{equation}
Here, $\Delta x$ and $\Delta z$ are the minimum resolvable distance between the two scatterers in the transverse and axial dimension, respectively; $c$ is the speed of light; $z$ is the distance of the point scatterers from the visible surface; and the scanning area has a size of $2w \times 2w$~m$^2$. These resolution limits were derived for the confocal scanning configuration~\cite{o2018confocal}. For non-confocal scanning configurations, the transverse resolution theoretically decreases by a factor of two. Other works have also used signal processing techniques~\cite{10TOFSMIT2016}, linear systems approaches~\cite{pediredla2017linear}, or feature visibility~\cite{liu2019analysis} to bound localization and photometric error in NLOS imaging scenarios.

{\bf{ {Other NLOS imaging approaches.}}}

\begin{table*}[!t]
\begin{tabular}{|l|l|l|l|}
\hline
\multicolumn{1}{|c|}{\textbf{\begin{tabular}[c]{@{}c@{}}reconstruction\\ method\end{tabular}}} & \multicolumn{1}{c|}{\textbf{light source}} & \multicolumn{1}{c|}{\textbf{detector}} & \multicolumn{1}{c|}{\textbf{Example refs}}             \\ \hline
detection or localisation                                                                         & High-rep./single-shot laser             & SPAD or APD                            & \cite{me2,brooks}                     \\ \hline
backpropagation                                                                                & High rep. laser                 & streak camera, SPAD array              & \cite{velten,buttafava2015}           \\ \hline
light-cone transform, f-K migration                                                                        & High rep. laser                 & SPAD array                             & \cite{o2018confocal,Lindell:2019:Wave} \\ \hline
virtual/phasor  field                                                                              & High rep.  laser     & SPAD or SPAD Array                                   & \cite{liu_non-line--sight_2019,teichman2019,reza2019experimemts_phasor,Dove_19}     \\ \hline
Steady-state, occlusions, coherence                                                                               & CW laser, ambient light     & standard CMOS camera, APD                                    & \cite{11LW2016,saunders2019computational,torralba2012,Ori}     \\ \hline
machine learning                                                                              & pulsed or CW laser     & SPADs, standard CMOS camera                                    & \cite{Tancik:18,chen2019steady,pier,Metzler:20}     \\ \hline
\end{tabular}
\caption{{\bf{Summary table of methods and requirements}}, illustrating the main reconstruction techniques and the corresponding hardware requirements (based on current experimental implementations), providing an overview of the variety of opportunities and hardware options. More details and references are provided in the main text. \label{table}}
\end{table*}

It is worth mentioning that there are other techniques that do not require transient light imaging capability.\\
{\emph{Steady state}} systems use a continuous spatially confined light source and a slow, conventional camera or detector to detect spatial variations in the return light. In these systems, integration times of the detector are long enough to consider the time of flight of the light infinite and what is detected is always a steady state scene response. Klein et al. for example show that the location of a single hidden object can be estimated when using a Short Wave Infrared (SWIR) light source and camera \cite{11LW2016}. An intriguing modification of the steady state approach is to use occlusions in the scene, such as edges, to provide additional spatial information and to rely on motion and differential measurements to eliminate problems with background light. In suitable scenes these methods can provide detailed information about objects in the scene using inexpensive, passive visible light cameras and natural ambient light sources
\cite{torralba2012,bouman2017,baradad2018,boger2018,saunders2019computational,thrampoulidis2018exploiting,chen2019steady}.\\
{\emph{Interferometric}} approaches illuminate the scene with a coherent light source and analyze interference patterns in the returned light. For example, the spatial speckle of the returned light can be collected and analyzed to reconstruct 2D NLOS images~\cite{Ori}. This method makes use of the memory effect that preserves angular information in the interaction with thin scatterers. This limits existing demonstrations to imaging very small objects, covering a solid angle of no larger than several degrees when viewed from the wall. This could likely be improved by incorporating more information, such as speckle patterns from multiple coherent light sources. Spatial correlations within the reflected light from an observation wall can also be used to directly retrieve information of a hidden scene, made up e.g. of active, yet incoherent light sources \cite{dogariu}. Extending this concept to the temporal domain, i.e. tracking of the temporal correlations within the reflected beam, enables a time-of-flight approach with  an impressive 10 fs resolution \cite{boger2018}. We have also already mentioned adaptive shaping of the illuminating laser beam that can transform the wall into a mirror by using an input spatial phase on the beam that compensates for scattering from the first surface. This then allows to scan a focused spot across the scene and retrieve image information from the reflected light intensity during the scan \cite{ilya}.   {Finally, deep learning techniques have recently been demonstrated to provide a useful framework to solve challenging inverse correlography problems arising in interferrometric NLOS approaches~\cite{Metzler:20}.} 

%
%
Another group of interferometric methods is based on illuminating the hidden scene with a pulsed coherent source via the relay surface and interfering the returning light with a delayed local oscillator light beam derived from the same coherent illumination source. This can be thought of as a coherent time gating method that produces data that can be treated similarly to data from other time resolved detectors. An example of this used a setup similar to a time domain Optical Coherence Tomography (OCT) system~\cite{OCT2}. Interference is used here as a coherence gate to determine the time of flight of the light through the scene. The need for an adjustable optical delay line complicates this setup. Another approach  presented by Willomitzer et. al. uses interference between the speckle patterns created by a NLOS object  and the reconstruction is obtained by combining the results from different illumination frequencies. This has the same effect as using a short pulse but eliminates the need for a delay line. Other efforts into coherent NLOS imaging include speckle interferometry to detect motion~\cite{willomitzer2019,rangarajan2019}.



%
{\bf{Conclusions and future directions.}}
\begin{figure}[!t]
\centering
\includegraphics[width=8.5cm]{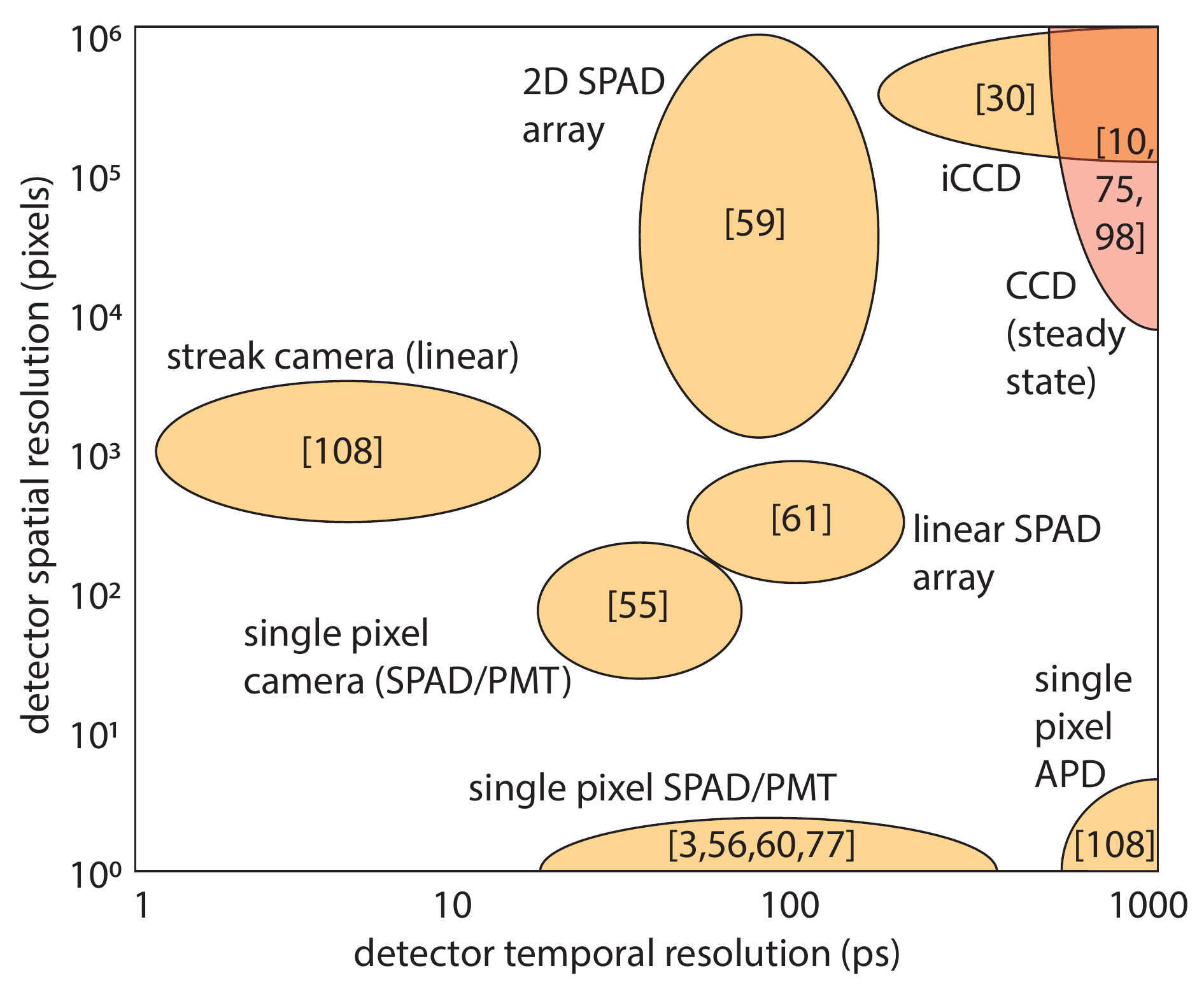} 
\caption{Summary of main detector technologies classified based on spatial and temporal resolution. Axes are n logarithmic scale. Steady state (i.e. using non-time-resolving) detector technologies such as CCD (or CMOS) cameras are also shown with a different colour in order to differentiate these from time-of-flight technologies. Some detector formats are linear, i.e. one-dimensional, as indicated in the graph. References in square brackets indicate example uses of each technology.}
\label{fig:summary}
\end{figure}

Light detection and ranging LiDAR systems are emerging as a standard imaging modality in autonomous driving, robotics, remote sensing, and defense. The same detectors---avalanche or single-photon avalanche photodiodes (APDs/SPADs)---are also increasingly used in consumer electronics, fluorescence lifetime microscopy, and positron emission tomography. 
SPADs in particular are an ideal platform for extending LiDAR to NLOS imaging because they address two of the primary challenges: being able to detect a few, indirectly reflected photons among many and time-stamping the photon time of arrival with high accuracy.

The capability to image objects outside the direct line of sight is likely going to be most useful for applications that already use LiDAR systems. For example, self-driving cars could sense obstacles beyond the next bend or in front of the car ahead of them could more safely navigate around them. Eventually, NLOS imaging could become a software-upgrade in existing or future LiDAR systems. For this and other reasons, we believe that such time-resolved NLOS imaging systems are one of the most promising directions in this emerging research area.

We have shown that there are multiple approaches and options for NLOS imaging, even when  restricted to time-of-flight techniques.  {In table~\ref{table}  and Fig.~\ref{fig:summary}, we summarise just the main techniques discussed here, together with the hardware requirements (based on current implementations)}. The nature of NLOS imaging is such that one should bear in mind that each approach has its own advantages and these need to be weighed-in when considering the specific application. For example, some NLOS LIDAR applications for the automotive industry may not actually require full 3D reconstruction of a scene but instead will benefit from a much simpler approach geared towards locating the position of a hidden object and identifying its nature (human, car, bicycle etc.). {Compared to alternative methods to image occluded spaces, such as transmitted or reflected RADAR, X-Ray transmission, reflected acoustic imaging or the placement of mobile cameras or mirrors,  optical NLOS imaging  {has the potential} to work in real time,  {in particular when considering large stand-off distances, albeit with targets that are limited to 2-3 meters behind the obstacle.} This kind of task becomes even more favourable when the hidden object is in movement as this allows easy subtraction of the background and has already been demonstrated to work at stand-off distances of 50 m or more in daylight conditions and recent reports indicate stand-off distances of 1.4 km. 
Simple range-finding from behind an obstacle can also be achieved with a single shot measurement if APDs instead of SPADs are used as these allow to collect multiple photons from a single, high energy return signal~\cite{brooks}. 

On the other hand, there are scenarios where full 3D imaging is indeed desired, for example in reconnaissance missions or in situations where 3D information of an otherwise inaccessible area is needed. Examples encountered by the authors range from identification of suitable underground cave sites for future manned planet missions to decommissioning of radioactive nuclear fission test facilities. In these situations the longer acquisition times required for 3D reconstruction and high laser powers may be less of an issue and an acceptable compromise. 
Existing systems could potentially already successfully tackle such tasks.

Finally, we mention that there is also scope for future work looking at combining the optical NLOS imaging techniques outlined here with other technologies. For example, radar and wifi can provide partial information directly through certain kinds of wall \cite{radar1,radar_walls,16RMTAC,15RTW,ralston_real-time_2010,Katabi1} and approaches based on sound \cite{Lindell:2019:Acoustic} have also been proposed. These and other technologies could potentially  complement each other either through direct data fusion so as to increase the collected information or alternatively by using one approach to provide course-grained information that would indicate the interest or need to continue with fine-grained optical techniques.


%
 


\bibliographystyle{unsrt}
\bibliography{Bib_for_Nature_review}

\end{document}